\documentclass{ws-ijmpc}

\usepackage{graphics}
\usepackage[dvips]{epsfig}
\usepackage[english]{babel}
\usepackage[latin1]{inputenc}

\begin{document}

\markboth{Ritter, Odenheimer, Heermann, Bantignies, Grimaud, Cavalli}
{Modelling and simulation of polycomb-dependent chromosomal interactions in drosophila}


\title{Modelling and simulation of polycomb-dependent chromosomal interactions in drosophila}

\author{Silke Ritter, Jens Odenheimer and Dieter W. Heermann}

\address{Institut f\"ur Theoretische Physik, Universit\"at Heidelberg, Philosophenweg 19,\\
D-69120 Heidelberg, Germany\\
j.odenheimer@tphys.uni-heidelberg.de\\
heermann@tphys.uni-heidelberg.de
}

\author{Frederic Bantignies, Charlotte Grimaud and Giacomo Cavalli}

\address{CNRS Institute of Human Genetics, 141, rue de la Cardonille\\
F-34396 Montpellier, France\\
Frederic.BANTIGNIES@igh.cnrs.fr\\
Giacomo.Cavalli@igh.cnrs.fr
}

\maketitle

\begin{history}
\received{Day Month Year}
\revised{Day Month Year}
\end{history}

\begin{abstract}
The conditions of the chromosomes inside the nucleus in the Rabl configuration have been 
modelled as self-avoiding polymer chains under restraining conditions. 
To ensure that the chromosomes remain stretched out and lined up, we fixed their end 
points to two opposing walls. The numbers of segments $N$, the distances $d_1$ and $d_2$ 
between the fixpoints, and the wall-to-wall distance $z$ (as measured in segment lengths) 
determine an approximate value for the Kuhn segment length $k_l$.
We have simulated the movement of the chromosomes using molecular
dynamics to obtain the expected 
distance distribution between the genetic loci in the absence of further attractive or repulsive forces. 
A comparison to biological experiments on \textit{Drosophila Melanogaster} yields information on 
the parameters for our model. With the correct parameters it is possible to draw conclusions on 
the strength and range of the attraction that leads to pairing.
\end{abstract}


\ccode{PACS Nos.: 87.15.Aa, 87.16.Sr, 87.14.Gg}

\section{Introduction}

The process of gene silencing is a crucial building block for the picture geneticists developed during the last decade about the functioning of genes within all kinds of organisms. Gene silencing is a highly complex area of research, and several mechanisms have been identified that inhibit gene expression within the nucleus.

The simplest molecular model to explain gene silencing postulates that specific repressors regulate the onset of transcription, by binding directly to specific DNA sequences and counteracting the action of activators and of the transcriptional machinery. A second possibility is that repressors, bound at specific sequences called silencing elements, might act by regulating the structure of the folded state of the DNA, called chromatin. In the cell nucleus, DNA is wrapped around histones to form the fundamental chromatin unit called nucleosome, and adjacent nucleosomes are able to fold into a higher-order chromatin fiber. These structures reduce the accessibility of DNA to the transcriptional machinery, and repressors might prevent transcription by stabilizing the binding of histones to DNA, or the folding of nucleosomes in compact higher-order chromatin structures.

Beside these levels of regulation, another level exists: namely the three-dimensional organization of chromosomal domains in the cell nucleus during cellular differentiation and development~\cite{Spector,Taddei}. In several cases, gene silencing has been correlated with relocation of chromosomal domains. In most of the published studies, gene silencing correlated with gene positioning close to heterochromatic compartments. Heterochromatin represents a highly compact region of chromatin where genes are stably repressed. 

\begin{figure}
\begin{center}
\includegraphics[width=0.7\textwidth]{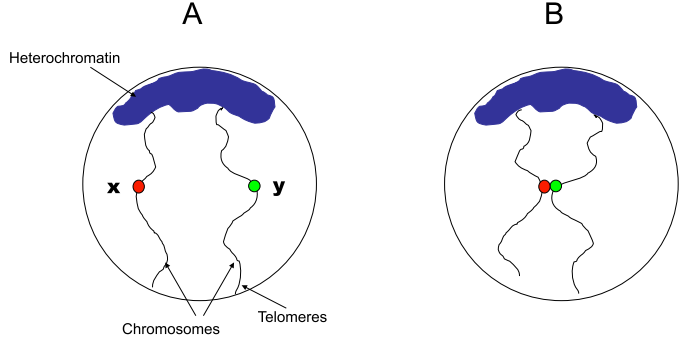}
\end{center}
\caption{Schematic representation of long-distance pairing between two loci X and Y in interphase nuclei. Nuclei are represented  in a Rabl configuration, in which centromeres are assembled near the apical pole of the nucleus, whereas the telomeres point toward the basal pole. The wavy lines represent the chromosomes projecting from centromeric heterochromatin toward the basal pole of the nucleus. Colored dots represent independent loci that are distant in a normal situation (A) but can come in close proximity upon integration of PRE-containing elements (B). This phenomenon leads to enhanced silencing and is dependent on PcG proteins.}
\label{fig:ban0307}
\end{figure}

Another case of gene silencing that shares common features with heterochromatin silencing involves the proteins of the Polycomb group (PcG)~\cite{Bantignies}. PcG proteins are highly conserved regulatory factors that are responsible for the maintenance of the silent state of important developmental genes, such as homeotic genes. In Drosophila melanogaster, PcG proteins form multimeric complexes and regulate their genes through binding to chromosomal regulatory elements named PcG response elements (PREs). This silencing involves repressive modifications on the target chromatin. In addition, it has been observed that silencing via PcG proteins and PREs is enhanced by the presence of multiple copies of PRE-containing elements in the nucleus. These copies may, but do not have to be on the same chromosome. Long-distance pairing between these two loci, which brings them closer together than they would usually be, leads to strong repression of the genes they control (Bantignies et al~\cite{Bantignies}). This type of regulation represents silencing by geometrical closeness, established in interphase nuclei  (see Fig.~\ref{fig:ban0307}).

In this report, we tried to model long-distance interactions among PREs, with the long term goal to build predictive models for proximity and interactions of chromosomal domains. Within the model, chromosomes were assigned a Rabl configuration~\cite{Rabl}  in the nucleus, a situation which is present in Drosophila embryonic nuclei. We calculated the expected distance distribution of the two loci in question and compared this distribution to experimental results that were obtained previously.

\section{The model}

\begin{figure}
\begin{center}
\includegraphics[width=0.8\textwidth]{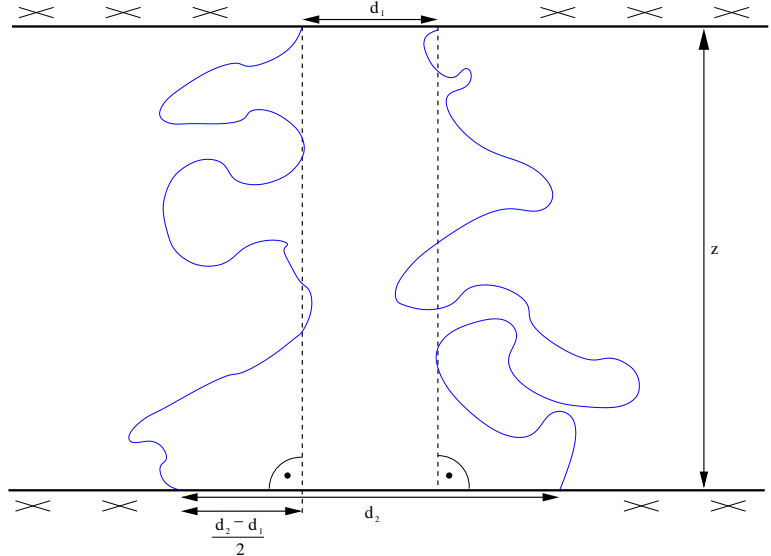}
\end{center}
\caption{Definition of the geometric simulation parameters. This
        sketch illustrates the meaning of the simulation parameters given in
        table~\ref{tab:parameters}. The chain contour length is $L=N\times a$.}
\label{fig:paramdef}
\end{figure}

In our model the two arms of the chromosomes carrying the gene loci are represented by 
two polymer chains. These chains are built up of ellipsoidal monomer segments with a ratio of $\frac{1}{20}$ between the half-axis (for a detailed description of the physical model used see~\cite{Odenheimer}). The chains have to be non-ideal, because the chromosomes cannot penetrate one another. Hence we have to demand that neither of the two monomers can occupy the same space at the same time. To ensure that our model is comparable to a Rabl configuration, where the DNA is ''stretched'' from the apical pole of the cell nucleus to its bottom pole (see Fig.~\ref{fig:ban0307}), we need a force acting on polymers which hinders them from curling. The easiest way to do so is to fix the chains between two walls. The distance between these walls should be large enough to ensure the stretching, yet small enough to allow for  movement of the chains. In our model the ratio between chain length and wall-to-wall-distance ranges 
from 1.5 to 5. Both chains have the same length, with the end-points fixed in a plane. The upper and lower end-points are symmetrical (see Fig.~\ref{fig:paramdef}).

The distance between the fixed points on the first wall ($d_{1}$) is kept small so they are essentially at the same point compared to the chain length. In our simulations we varied $d_{1}$ between 1 and $\frac{1}{4}$ monomer length. 

\begin{table}
\caption{Simulation parameters. The values for the different distances characterizing restraining conditions. Here a refers to the length of one monomer/one step in the self-avoiding random walk (SAW).}
\begin{center}
\begin{tabular}{c|c|c|c|c}
  \hline
  \hline
  Number of  & Wall-to-wall  & Distance of  & Distance of  & App. number of  \\
  monomers & distance & fixpoints on & fixpoints on & configurations \\
  in one chain $N$ & $z$ $[a]$ & $1^{st}$ wall $d_{1}$ [a]& $2^{nd}$ wall $d_{2}$ $[a]$& obtained up to date\\
  \hline \hline
  \multicolumn{5}{c}{\large Parameter Set \textbf{A}}\\
  60 & 30 & 0.25 & 0.25 & 13,000\\
  80 & 30 & 0.25 & 0.25 & 31,000\\
  100 & 30 & 0.25 & 0.25 & 13,500\\
  150 & 30 & 0.25 & 0.25 & 10,000\\
  \hline \hline
  \multicolumn{5}{c}{\large Parameter Set \textbf{B}}\\
  60 & 30 & 0.25 & 30 & 14,800\\
  80 & 30 & 0.25 & 30 & 25,900\\
  100 & 30 & 0.25 & 30 & 28,100\\
  150 & 30 & 0.25 & 30 & 4,000\\
  \hline \hline
  \multicolumn{5}{c}{\large Parameter Set \textbf{C}}\\
  60 & 45 & 0.25 & 0.25 & 14,000\\
  80 & 45 & 0.25 & 0.25 & 31,000\\
  100 & 45 & 0.25 & 0.25 & 14,200\\
  150 & 45 & 0.25 & 0.25 & 13,800\\
  \hline \hline
  \multicolumn{5}{c}{\large Parameter Set \textbf{D}}\\
  60 & 30 & 1 & 1 & 18,000\\
  80 & 30 & 1 & 1 & 8,500\\
  100 & 30 & 1 & 1 & 12,200\\
  150 & 30 & 1 & 1 & 8,500\\
  \hline \hline
\end{tabular}
\end{center}\label{tab:parameters}
\end{table}

For a free SAW some essential properties are well known (see e.g. \cite{gen79}). Denoting the end-to-end distance by $\vec{r_{e}}$, its mean by $\bar{r_{e}}=<\mid \vec{r_{e}} \mid>$ and its root mean square by $R_{E}=\sqrt{<(\mid\vec{r_{e}}\mid-\bar{r_{e}})^{2}>}$, then for long walks (number of steps $N$ large), we know that $R_{E}$ scales as:
\begin{equation} \label{eqn:refree}
R_{E}(N) \approx aN^{\nu}
\end{equation}
$\nu$ is called the Flory parameter, initially calculated by Flory~\cite{flo71} to $\frac{3}{5}$, which is still a very good approximation.
When looking at the end-to-end distance distribution function $p(\vec{r_{e}})$ for a free SAW, there are two different scaling laws, one for the region where $\mid \vec{r_{e}} \mid$ is small and one for the region where it is large.
For small values the distribution follows a power law:
\begin{equation}
\lim_{\frac{r_{e}}{R_{E}} \rightarrow 0}p(\vec{r_{e}})=\textrm{constant}\times\left(\frac{r_{e}}{R_{E}}\right)^{\frac{\gamma-1}{\nu}}
\end{equation}
whereas for large values a stretched exponential dampening factor dominates:
\begin{equation}
\lim_{\frac{r_{e}}{R_{E}} \rightarrow \infty}p(\vec{r_{e}})=e^{-(\frac{r_{e}}{R_{E}})^{\frac{1}{1-\nu}}}\times f_{1}\left(\frac{r_{e}}{R_{E}}\right), 
\end{equation}
\label{eqn:upperscale}
where $f_{1}$ is some polynomial function.
For the free case, $\gamma$ is a universal parameter that depends solely on the dimensionality $d$. For $d$=3 we have $\gamma\approx \frac{7}{6}$. The distribution $p(r_{e})$ depends on $r_{e}$ solely through the ratio $\frac{r_{e}}{R_{E}}$.
Changes in the geometry of the chain environment (e.g. by the introduction of wedges or walls) lead to changes in the value of the parameter $\gamma$, but leave the overall scaling predictions unaffected~\cite{car84}. In our study we are simulating two stretched chains with fixed endpoints, which might be considered as a single one as long as the fixpoint distance $d_{1}$ is kept small. One of our goals is to determine the shape of the distance distributions between arbitrary monomers under these highly restraining conditions.\\
Even if considering the distribution function of the distance $\vec{r_{i,j}}$ between two arbitrary steps $i$ and $j$ of a free SAW, the form of the distance distribution changes from the end-to-end case. Though we might expect that a section of a free SAW would behave also like a shorter free SAW itself, this is not the case. The loss of entropy at the endpoints leads to a loss of the shape properties.\\
In our case, the chains are fixed between two walls and therefore a force
is exerted on them which restricts their ability of movement.
Additionally, we have to expect further changes in the properties of the chains due to loss of entropy as we fix the position of the end point vectors of the walk precisely (the wall fixpoints), and others have to
stay in restricted areas (because both start and end points of the walk
are fixed). This means we are sampling over a distinct subunit of the
ensemble of configurations of all free SAWs, which most probably has
consequences on the distributions $p_{n_{1},n_{2}}(r_{n_{1},n_{2}})$ of the distances between monomer $n_{1}$ on chain 1 and monomer $n_{2}$ on chain 2. Determining the shape of these distance distributions has been an important aspect of our simulations. We were able to verify by $\chi^{2}$-minimum fits to ascertain that for most combinations of ($n_{1}$,$n_{2}$), $p_{n_{1},n_{2}}$ is Gaussian. However, in those situations where the distribution had a small mean value, the shape clearly deviated from the Gaussian form. In these cases our fits pointed towards a stretched exponential for the upper end of the distribution, where the stretching exponent is different from $\frac{1}{1-\nu}$ which was valid for the free end-to-end distance distributions (see equation~\ref{eqn:upperscale}). This observation will have to be verified on simulations with improved statistics.

\section{Comparison with experiment}

Our simulations will be compared with two different sets of experiments. In both cases the two genetic loci in question were positioned on different chromosomes. This makes our model applicable as there is no direct connection between the chromosomal arms we consider. If they were positioned on different arms of the same chromosome, the centromer region might serve as a transmitter, any ''tugging motion'' of one locus might be perceptible for the other one. Fixing both chains to different fixpoints on the wall does not incorporate this feature.
In both experiments, copies of an element called {\it Fab-7} ({\it Fab-7} is an element containing a PRE that regulates an homeotic gene of Drosophila melanogaster) have been inserted into different loci X and Y:

\begin{description}
\item[Experiment I] used \textit{Drosophila} embryo cells, for which {\it Fab-7} had been inserted at locus X1 (``{\it sd}'') and Y1 (``BX-C''). These loci are situated approximately at one third from the top of the nucleus according to F.~Bantignies.
\item[Experiment II] used \textit{Drosophila} embryo cells, for which {\it Fab-7} had been inserted at locus X2 (``{\it sd}'') and Y2 (``38F''). Locus X2 (``{\it sd}'') is still approximately one third from the top of the nucleus, but locus Y2 (``38F'') is much closer to it ($\approx$ one sixth to one eighth) according to F.~Bantignies.
\end{description}

\begin{figure}
\begin{center}
\includegraphics[width=0.5\textwidth]{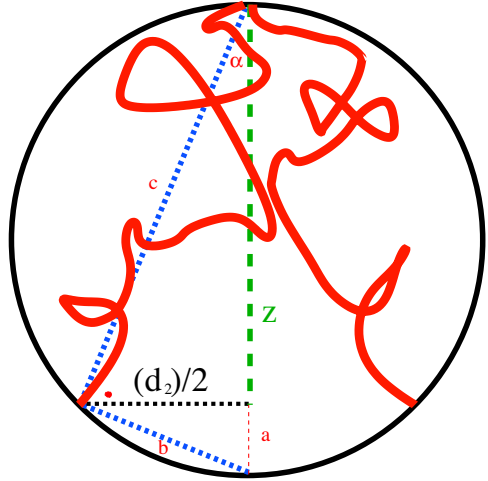}
\end{center}
\caption{Determination of length scale. Here $z$ is the wall-to-wall distance of our simulation setup, $d_{2}$ is the distance between the fixpoints on one wall (the distance $d_{1}$ on the other wall has been neglected as $\approx 0$). The angle $\alpha$ as well as the distances a, b and c have been introduced solely for the sake of calculating the conversion factor from arbitrary length scale to micrometer and have no further meaning.}
\label{fig:spherescale}
\end{figure}

To make the distance distributions obtained through the simulations comparable to these experimental results, we have to scale the data. We may assume that the fixpoints on the walls in our simulation correspond to (loose) binding of the chromosomes to the membrane of the nucleus as outlined in Fig.~\ref{fig:spherescale}. 
We neglected the fix point distance at the top ($d_{1}$) as it is quite small compared to $z$. 
As we know the diameter $D$ of the sphere, we can use it to find an approximate scaling relation. In fact, this is the highest scaling factor we may assume that allows us to map the simulation setup into the inside of the nucleus. If we were to consider the less likely scenario where the ends of our chromosomes are not fixed to the cell membrane, but to some other place within the nucleus, our scaling factor would have to be chosen smaller, as the distance between the beginning and the end points of the chains, which we hold fixed, will correspond to less than the diameter of the nucleus. Any larger Kuhn segment length would place at least one of the fixpoints outside the nucleus.

The length $c$ in Fig.~\ref{fig:spherescale} is most suitable for the derivation. In units of $D$ we have:
\begin{eqnarray}
\tan\alpha&=&\frac{d_{2}}{2z} \nonumber\\
c[D]&=& D \cos \alpha,
\end{eqnarray}
whereas in our arbitrary length scale, whose units we shall call $[a]$, we have:
\begin{equation}
c[a]=\sqrt{z[a]^{2}+\frac{d_{2}[a]^{2}}{4}}
\end{equation}
which, using $D\approx5 \mu m$, leads to
\begin{equation}
1 [a] \approx \frac{D\cos(\tan^{-1}(\frac{d_{2}}{2z}))}{\frac{1}{[a]}\sqrt{z[a]^{2}+\frac{d_{2}[a]^{2}}{4}}} [\mu m].\label{eqn:atomicrom}
\end{equation}
For the parameter sets A, C and D (see table~\ref{tab:parameters}), where $d_{2}$ is much smaller than the wall-to-wall distance $z$, we may approximate equation~(\ref{eqn:atomicrom}) so that $1[a] \approx \frac{5}{z}[\mu m]$, in case B we have to use the complete formula.\\
The value of the Kuhn segment length for chromosomes is not yet well known, though most biologists agree that it is of the order of a few hundred $nm$ (see e.g.~\cite{rip01}).
When we rescale our data according to equation~(\ref{eqn:atomicrom}), we also determine the resultant Kuhn segment and chain contour length (see table~\ref{tab:chainlengthsinmicrom}). The Kuhn segment lengths are in the right order of magnitude. To have an immediate comparison to the experimental histograms, we also performed a rebinning of our simulation data into the same ranges Bantignies et al. used in their data processing.

\begin{table}
\caption{Effective chain and Kuhn segment length for the different parameter sets. The geometry of the system fixed the scaling relations between arbitrary units of the Kuhn segment length to $\mu m$.}
\begin{center}
\begin{tabular}{c|c|c}
  \hline
  \hline
  Number of segments & Approximate chain & Resultant effective Kuhn\\
  per chain $N$ & length $L$ [$\mu m$] & segment length $l_{k}$ [$nm$]\\
  \hline \hline
  \multicolumn{3}{c}{\large Parameter Set \textbf{A}}\\
  60 & 10 & 170 \\
  80 & 13 & 170 \\
  100 & 17 & 170 \\
  150 & 25 & 170 \\
  \hline \hline
  \multicolumn{3}{c}{\large Parameter Set \textbf{B}}\\
  60 & 8 & 130 \\
  80 & 11 & 130 \\
  100 & 13 & 130 \\
  150 & 20 & 130 \\
  \hline \hline
  \multicolumn{3}{c}{\large Parameter Set \textbf{C}}\\
  60 & 6 & 110 \\
  80 & 9 & 110 \\
  100 & 11 & 110 \\
  150 & 16 & 110 \\
   \hline \hline
  \multicolumn{3}{c}{\large Parameter Set \textbf{D}}\\
  60 & 10 & 170 \\
  80 & 13 & 170 \\
  100 & 17 & 170 \\
  150 & 25 & 170 \\
   \hline
  \hline
\end{tabular}
\end{center}\label{tab:chainlengthsinmicrom}
\end{table}

\begin{figure}
\begin{center}
\includegraphics[width=0.8\textwidth, angle=270]{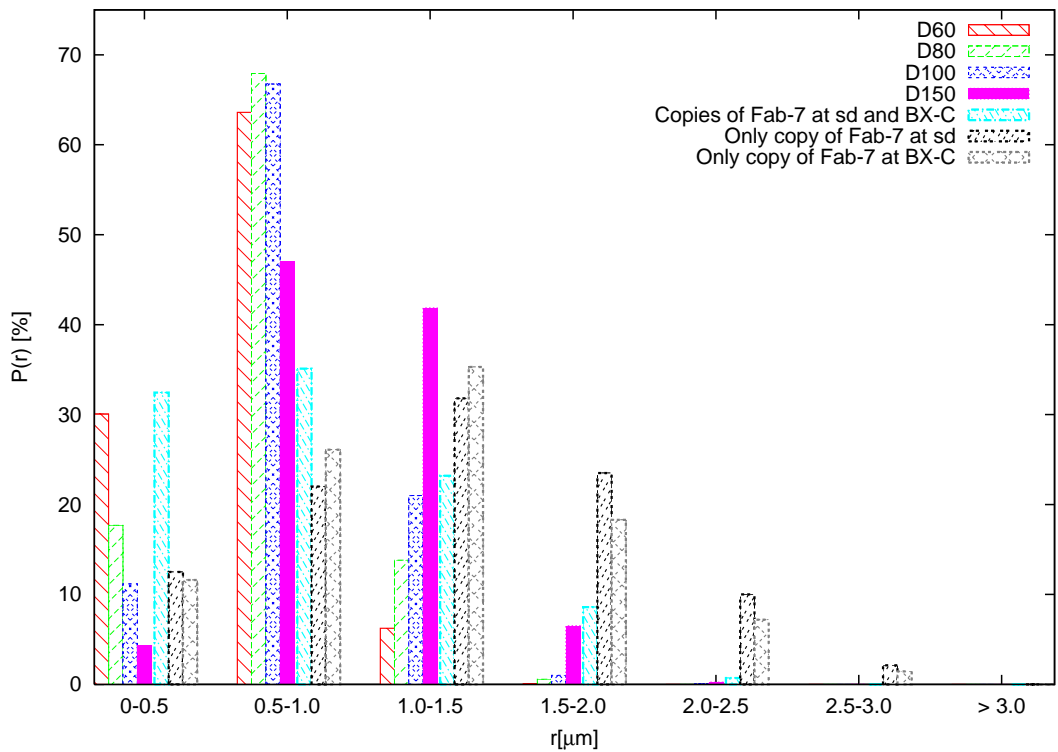}
\end{center}
\caption{Here we see the different distance distributions that involve the {\it sd} and the BX-C loci. 
The four bars on the left hand side in each group represent different chain lengths of 60, 80, 100, and 150 Kuhn segment
lengths, respectively.  In the case shown the Kuhn segment length was 170 nm. The next bar (turquoise) gives the experimental results when two copies of {\it Fab-7} are present,
the black and grey bars stand for the control groups with only one copy.}
\label{fig:ExpComp}
\end{figure}

\begin{figure} 
\begin{center} 
\includegraphics[width=0.8\textwidth, angle=270]{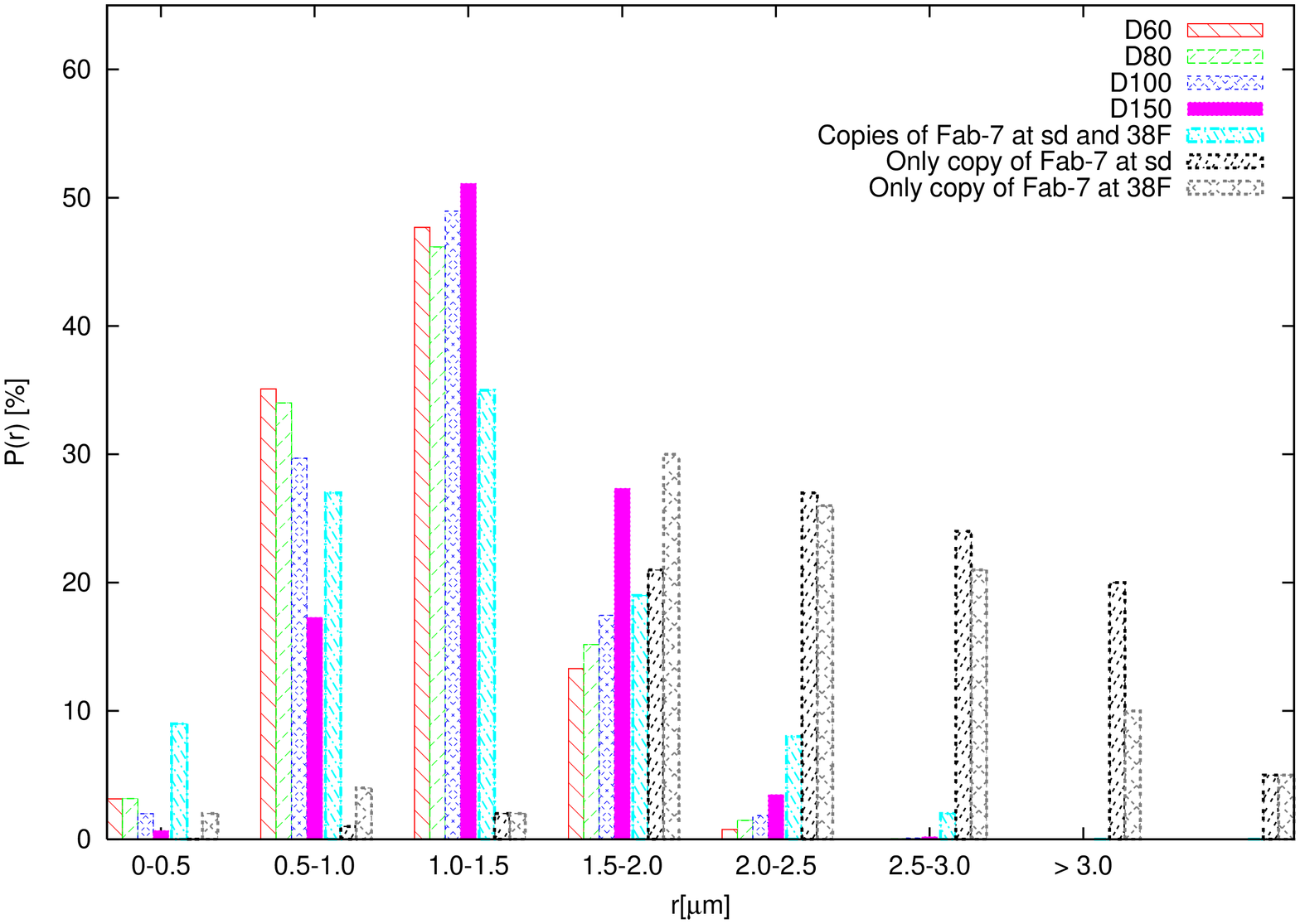} 
\end{center} 
\caption{In the above graphics we see the different distance
distributions that involve the {\it sd} and 38F loci. The legend is identical to Fig.~\ref{fig:ExpComp}.} 
\label{fig:ExpComp2} 
\end{figure}

We evaluated the distance distributions between point $n_{1}$ on chain 1 and point $n_{2}$ on chain 2. In accordance with the position of the genes along their chromosomes,
for experiment I ({\it sd} and BX-C) we chose $n_{1},n_{2} \in[0.3N;0.4N]$ , where $n=0$ would correspond to the fixpoint on the top wall. For experiment II ({\it sd} and 38F) we decided to consider ranges of $n_{1} \in [0.1N;0.2N]$ and $n_{2} \in [0.3N;0.4N]$. An example of the resulting graphs for parameter set D can be seen in Figs.~\ref{fig:ExpComp} 
and~\ref{fig:ExpComp2}. By
visual judgement of these graphs we compared the simulation data to the biological experimental data. When checking for possible agreement between the model and the
experiment, we realized that there is most probably the possibility of identifying the distance distributions in the groups with two copies of the gene with distributions
from our simulation. We realized that congruence between the curves increased as the number of segments per chain increased. We also found that a larger Kuhn segment length
is favorable. In these comparisons we always neglected discrepancies in the very first bin (smallest distances) as the biological model predicts a short-range binding force
keeping the genes together once they come within range. This force has not yet been incorporated in the simulation model. On the other hand, the graphs also give strong
evidence that no parameter choice within our simple model allows for identification between the control group distributions and the simulation, because the full-width
half-maximum (FWHM) of the experimental curves is large although the mean value is relatively small. Parameter choices that could produce a comparable FWHM in the simulation
would always place the mean value of the distribution at a much larger value.

\section{Conclusions}

The simple ''two chains, no interaction'' (except self-avoidance) model employed in the simulations is most likely adequate to account for the observed distance distributions in the case of presence of two copies of {\it Fab-7} (the ''pairing'' case).
We were able to distinguish two trends: Those simulations that had the most segments were best suited to ''reproduce'' the experimental curve, as were setups with increasing Kuhn segment length. Using this information we have a good idea in which areas of the parameter space further simulations should be conducted to enable a numerical comparison between experimental and simulation data with the aim of determining strength and range of the pairing/binding force between the two copies of {\it Fab-7}.

The failure of the model with regard to the control groups is rooted in its simplicity. Possibly there is a repulsive force acting that is keeping different genes in different compartments of the nucleus. The simulation data we already have together with further biological data from Bantignies et al. concerning the distances not only between the two genetic loci, but also between other places along the chromosomes in question should enable us to improve our model by introducing adequate attractive and/or repulsive forces.

An open question concerns the shape of the distance distribution function in different regimes. Whereas in some cases, $\chi^{2}$-minimum fits yield that the distribution is clearly Gaussian, in others notable deviations were found, especially in distributions with small mean distances. Though we found evidence that the upper end of these might behave as a stretched exponential, further investigations with improved statistics will have to be conducted to solidify this result. 

\section{Acknowledgment}

J.O. is supported by the DFG Project KR 2213/2-1.
F.B. is supported by the CNRS. G.C. was supported by grants of the CNRS, the Human Frontier Science Program Organization, the European Union FP6 (Network of Excellence The Epigenome and STREP 3D Genome), by the Indo-French Centre for Promotion of Advanced Research, and by the Minist$\grave{e}$re de l'Enseignement Sup$\acute{e}$rieur, ACI BCMS. 


\end{document}